\documentclass[prd,nofootinbib]{revtex4}

\usepackage{epsfig}
\usepackage{graphicx}
\usepackage{amssymb,amsmath}
\usepackage{epstopdf}

\newcommand{\be}{\begin{equation}}
\newcommand{\ee}{\end{equation}}
\newcommand{\bea}{\begin{eqnarray}}
\newcommand{\eea}{\end{eqnarray}}
\renewcommand{\b}[1]{\overline{#1}}

\begin{document}

\title{Anomalous conductivity in holographic QCD}

\author{Gilad Lifschytz}
\affiliation{Department of Mathematics and Physics and CCMSC\\
University of Haifa at Oranim\\
Tivon 36006, Israel}
\email{giladl@research.haifa.ac.il}

\author{Matthew Lippert}
\affiliation{Department of Physics\\
Technion, Haifa 32000, Israel\\
{\rm and}\\
Department of Mathematics and Physics \\
University of Haifa at Oranim\\
Tivon 36006, Israel}
\email{matthewslippert@gmail.com}

%\date{}

\begin{abstract}
We compute the longitudinal and Hall conductivities in the parallel phase of the Sakai-Sugimoto model with a transverse magnetic field. We find that the conductivities behave as if the charge of the system is made out of two different types; one behaves as charge carriers flowing through a dissipative neutral medium, while the other does not feel the dissipation.  We also investigate the case of an electric field parallel to the magnetic field and find that in this case the system behaves as a perfect conductor.  Both of the unusual behaviors stem from the axial anomaly.
\end{abstract}

\maketitle

\section{Introduction}

%SS good for QCD but also for strongly coupled electrons
The Sakai-Sugimoto model is an attempt to study phenomena related to QCD through a holographic description which captures the properties of flavor fermions strongly interacting with gluonic degrees of freedom \cite{SS}.  However, one can alternately view this model as simply an interesting system of strongly interacting fermions in four dimensions which, as a consequence of holographic duality, is particularly tractable.

%review of recent results of SS in EM: pion gradient, axial current, smeared 4-branes
Recently there have been investigations \cite{BLL3, ST, RSS} of the Sakai-Sugimoto model with a background magnetic field and with nonzero baryon density which reproduced certain properties related to the axial anomaly.  In particular, in the confined phase the magnetic field generates, through the anomaly, a pion gradient which, as in QCD \cite{SonStephanov}, carries baryonic charge, and at large magnetic field the preferred phase is one in which all the baryon charge is carried by this pion gradient.  In the chiral-symmetric deconfined phase the magnetic field generates, again via the anomaly, an axial current \cite{MZ, SonNewman}.  Furthermore, at nonzero baryon density the application of a magnetic field causes the baryonic charge, which at zero field was holographically represented by sources either at the horizon or at the bottom of the U-shaped 8-brane, to rise up and become smeared along the radial direction of the 8-brane.  In the confined phase the treatment of a pion gradient phase without extra sources opens up the possibility of studying a phase with baryonic charges without the approximations needed to consider sources at the tip of the U-shape. 

%what we do here: add electric field, compute conductivity for orthogonal and parallel, 
In this letter we will add an electric field orthogonal to the magnetic field and study the properties of the conducting chiral-symmetric phase using both the Kubo formula \cite{SonStarinets, PhotonDilepton, PS2} and the Karch-O'Bannon method \cite{KOB, OB}.  We find an interesting structure; both the longitudinal and the Hall DC conductivities are consistent with the existence of two types of charge carriers.  One type has the response of particles interacting with a dissipative neutral media, and the other type has no dissipation.  They correspond in the bulk five-dimensional picture to charges at the horizon and charges spread radially along the 8-brane.

We further consider an the electric field parallel to the magnetic field.  In this case the anomaly results in the non-conservation of the axial charge which in turn generates a baryonic current.  We compute the conductivity in this case and show it is proportional to a delta function at zero frequency.  Thus, the finite temperature vacuum can be considered as a perfect conductor.
 
%%%%%%%%%%%%%

\section{Review of Holographic QCD with background fields}
We first present an abbreviated review of the Sakai-Sugimoto model with nonzero baryon density and with background electromagnetic fields.  For more detailed discussion see, for example, \cite{SS, ASY, PS1} for the model and finite temperature phase structure, \cite{BLL1, UCLA, UBC, KSZ1} for nonzero baryon density, and \cite{PS2, BLL2, USC, KSZ2} for electric and \cite{BLL3, ST, RSS} for magnetic fields.

% SS basic description - bulk, boundary, confinement, chiral symmetry breaking
The Sakai-Sugimoto model consists of $N_f$ 8-branes and $N_f$ $\b 8$-branes treated as probes in the near-horizon geometry of $N_c$ 4-branes wrapped on an $S^1$.  At large $g_sN_c$, this bulk description is dual to a four-dimensional nonsupersymmetric, strongly-coupled $SU(N_c)$ gauge theory with massless chiral fermions having flavor symmetry $U(N_f) \times U(N_f)$.  This gauge theory is confining at low temperature, but when the bulk undergoes a Hawking-Page transition and develops a horizon, the gauge theory correspondingly deconfines.  Chiral symmetry is determined by the embedding of the 8-branes.   When the 8-brane and $\b 8$-brane connect, as is required by topology in the confining phase, the $U(N_f) \times U(N_f)$ is broken to the diagonal $U(N_f)$, but if the branes remain disconnected, ending on the horizon in the deconfined phase, chiral symmetry is restored.

%set up 
We will be interested in the chiral-symmetric, deconfined phase because it is also the conducting phase and will consider the one-flavor case, $N_f=1$.  
%dimensionless variables
For convenience and clarity, we will work primarily in dimensionless variables which will be denoted by lowercase letters.  These are related to the physical uppercase dimensionful variables by appropriate powers of $\alpha'$ and the bulk radius of curvature $R$.  For example, the dimensionless coordinates are $x^\mu = X^\mu/R$ and $u = U/R$, and the 8-brane gauge field is $a_\mu = \frac{2\pi \alpha'}{R} A_\mu$.  A complete table of conversions is given in appendix A of \cite{BLL3}.

%metric, background fields, 8-brane embedding
The bulk metric for the deconfined phase is
\be
ds^2 = u^{3/2} \left(-f(u)dt^2 + d{\bf x}^2 + f(u) dx_4^2\right) + u^{-3/2}\left({du^2\over f(u)} + u^2 d\Omega_4^2\right) 
\ee
where $x_4$ is a compact coordinate with periodicity $2\pi R_4/R$, and 
\be
f(u) = 1- {u_{T}^3\over u^3} \; , \; u_{T} = \frac{16\pi^2 R^2 T^2}{9} \,.
\ee
The dilaton and RR four-form are 
\bea
e^{\Phi} &=& g_s u^{3/4} \\
F_4 &=& 3\pi (\alpha')^{3/2}  N_c \, d\Omega_4 \,.
\eea
The curvature radius of the space is given by
\be
R = (\pi g_s N_c)^{1/3} \sqrt{\alpha'} \; ,
\ee
and this is related to the four-dimensional 't Hooft coupling 
\be
\lambda = {4\pi g_s N_c \sqrt{\alpha'}\over R_4}\,.
\ee
In the conducting, chiral-symmetric phase, the 8-branes and $\b 8$-branes are parallel, at fixed positions in $x_4$, and extend from the boundary at $u=\infty$ to the horizon at $u=u_T$.  

% baryons
We would like to consider the system at nonzero baryon density.  The baryonic $U(1)_V$ symmetry of the gauge theory corresponds to the diagonal $U(1)_V$ of the 8-brane gauge field $a_\mu$, and the holographic dictionary relates the baryon current $j_\mu$ to the variation of the bulk on-shell action with respect to $a_\mu$:
\be
\label{currents}
j_\mu = \frac{\delta S \, \vline_{EOM}}{\delta a_\mu (\infty)}=  {\frac{\delta L}{\delta a'_\mu} \vline}_{u \to \infty}  \ .
\ee
From this we see that baryon density $d\equiv j_0$ is then given by the $u$-component of the electric displacement at infinity, and the baryon chemical potential is dual to the potential at infinity, $\mu = a_0(u=\infty)$.  
%sources-  quarks, baryons = D4
In the chiral-symmetric parallel phase, the 8-branes can carry a nonzero electric field without sources because they end on the horizon.  Here one can view $a_\mu$ as being sourced by 4-8 and 4-$\b 8$ strings which, in the near-horizon limit, are located at the intersection of the horizon and the 8-branes and which represent the chiral fermions of the gauge theory.  When the 8-branes connect and chiral-symmetry is broken, a nontrivial $a_\mu$ must arise from a source;  these are either strings stretching from the horizon which correspond to massive quarks or 4-branes wrapped on $S^4$ which are described by instantons in the world-volume of the 8-branes and correspond to gauge theory baryons.

%EM U(1) = baryon U(1)
While there is no true electromagnetic field in the Sakai-Sugimoto model, we can use the non-normalizable mode of $a_\mu$ to model one.  Since $a_\mu$ is dual to the baryon current, for $N_f=1$ electric charge is then just identified with baryonic charge.  We are using different components of the same gauge field $a_\mu$ to describe both the background field and the currents coupling to it:  the background electromagnetic field strength is $f_{\mu\nu} = \partial_{[\mu} a_{\nu]}$, independent of $u$, whereas the current is $j_\mu$ is related to $\partial_u a_{\mu}$.\footnote{We will work in the $a_u = 0$ gauge.}  Note that since there is no dynamical four-dimensional electromagnetic field, the charges and currents do not source the fields.  Therefore, there is no difference between the magnetic field $h$ and the magnetic induction $b$.

%background E - current, conductivity without B
We analyzed the effect of a background electric field $e$ in \cite{BLL2}.  In the deconfined phase, the chiral-symmetric parallel phase supports a current, and the chiral symmetry breaking transition is also a conductor-insulator transition.  In the bulk, strings stretching from the 8-branes to the horizon can flow along the horizon which corresponds in the gauge theory to flowing chiral fermions interacting with the gluons.  Using both the Karch-O'Bannon method \cite{KOB, OB} and the Kubo formula \cite{SonStarinets, PhotonDilepton, PS2}, we computed the conductivity which, to leading order in the electric field, is
\be
\label{conductivityh=0}
\sigma =  \sqrt{\frac{u_T^5+d^2}{u_T^3}}
\ee
We will see in this work how this formula is generalized with the addition of a magnetic field, and we will find an orthogonal Hall current as well.

%background B - anomaly, axial current, smeared 4-branes=pion gradient
In \cite{BLL3, ST} the effects of a background magnetic field $h$ at nonzero density, in particular those mediated by the axial anomaly, were investigated.  A magnetic field in the 1-direction leads to a non-trivial $a_1(u)$.  For the confined phase, this is dual to a pseudo-scalar gradient\footnote{Because we considered only one flavor, the pseudo-scalar here is the $\eta'$.}
\be
\nabla \varphi \propto \mu h \ ,
\ee
while in the chiral-symmetric deconfined phase, a magnetic field generates a parallel axial current 
\be
j_A = \frac{3}{2} \mu h \ .
\ee
Furthermore, $a_1(u)$ acts as an additional source of electric charge on the 8-brane.  We can identify the magnetic field $f_{23}$ as a smeared 6-brane in the 8-brane and the anomaly-generated $f_{u1}$ as an orthogonally smeared 6-brane.  The two crossed 6-branes in turn carry 4-brane charge.  Just as an instanton in the 8-brane gauge theory, which was identified as a wrapped 4-brane, was the holographic dual of a baryon, this magnetically induced, smeared 4-brane holographically carries baryon charge as well.  There are now two different bulk objects that carry baryon charge in the chiral-symmetric phase: short strings from the horizon to the 8-branes and 4-branes smeared inside the 8-branes.  With the addition of an electric field, we will see that these two types of charge carriers have remarkably different transport properties.

%%%%%%%%%%%%%%

\section{Orthogonal electric and magnetic fields}
\label{orthogonal}

\subsection{Conductivity by Karch-O'Bannon}
\label{KarchOBannon}

We will first consider orthogonal electric and magnetic fields and calculate the ohmic and Hall conductivities employing the same method used by Karch and O'Bannon in the D3-D7 model \cite{KOB, OB}.  We consider the deconfined, chiral-symmetric phase with background magnetic field $f_{23}=\partial_2 a_3 = h$ and electric field $f_{02} =\dot a_2 = e$.  In order to include spacetime baryonic currents $j_\mu$ (though we will call the baryon density $d \equiv j_0$), the spacetime components of the gauge fields will have $u$-dependent parts $a_\mu(u)$ as well.  Primes denote derivatives with respect to $u$ and dots are derivatives with respect to $t$.

We combine the actions of the 8-brane and the $\b 8$-brane by considering the vector part of $a_0$ and the axial part of $a_1$, i.e. $a_0^{\b {D8}} = a_0^{D8}$ and  $a_1^{\b {D8}} = - a_1^{D8}$.  The DBI action with these background fields is
\be
\label{DBI}
S_{\rm DBI} = {\cal N} \int du \, u^{5/2} \sqrt{X}
\ee
where
\be
X = 1+ \frac{h^2}{u^3} - \frac{e^2}{f(u)u^3} - {a_0'}^2\left(1+ \frac{h^2}{u^3}\right) +  f(u){a_1'}^2\left(1+ \frac{h^2}{u^3} - \frac{e^2}{f(u)u^3}\right) + f(u){a_2'}^2 + f(u){a_3'}^2\left(1 - \frac{e^2}{f(u)u^3}\right) - 2\frac{eh}{u^3} \, a_0' a_3' \ .
\ee
The normalization constant is given by
\be
{\cal N} = 2 \Omega_4 T_{D8} R^5 = \frac{N_c}{6\pi^2} \frac{R^2}{(2\pi\alpha')^3}
\ee
where the factor of 2 corresponds to the two halves of the embedding.  The Chern-Simons action is
\be
S_{\rm CS} =  {\cal N} \int du \, \left(h a_{[0} a_{1]}'  + e a_{[1}a_{3]}' + a_2 a_{[1}' \dot a_{3]} + a_3\partial_2 a_{[0} a_{1]}'  \right) \ .
\ee
To deal with the last two terms which give contributions at the boundary, we add to the action the following boundary terms (as in \cite{BLL3}):
\be
\Delta S = \frac{ {\cal N} }{2} \int \left\{ \partial_2\left(a_3 a_{[1} a_{0]}' \right) +  \partial_0\left(a_2 a_{[1} a_{3]}' \right) + \partial_u\left(a_3 \partial_2a_{[1} a_{0]} + a_2 \dot a_{[1} a_{3]} \right) \right\} \ .
\ee
The combined action simplifies to
\be
\label{CS}
S_{\rm CS} + \Delta S = \frac{3 {\cal N} }{2} \int \left(h a_{[0} a_{1]}'  + e a_{[1}a_{3]}' \right) \ .
\ee
The equations of motion derived from (\ref{DBI}) and (\ref{CS}) and then integrated with respect to $u$ are:
\bea
\label{EOM03}
\frac{u^{5/2} \left(1+\frac{h^2}{u^3}\right)a_0' + \frac{eh}{u^{1/2}} a_3'}{\sqrt{X}} &=& c_0 + 3 h a_1  \\
\label{EOM1}
\frac{u^{5/2}\left(1+\frac{h^2}{u^3} -  \frac{e^2}{f(u)u^3}\right) f(u)a_1'}{\sqrt{X}} &=&  c_1 + 3 h a_0 +  3 e a_3\\
\label{EOM2}
\frac{u^{5/2} f(u) a_2'}{\sqrt{X}} &=&c_2\\
\label{EOM30}
\frac{u^{5/2}\left(1- \frac{e^2}{f(u)u^3}\right) f(u)a_3' - \frac{eh}{u^{1/2}} a_0'}{\sqrt{X}} &=& c_3 - 3 e a_1 
\eea
where $c_\mu$ are constants of integration which can be written in terms of the baryonic currents. 

The baryonic currents are given holographically by the boundary values of the momenta conjugate to the gauge fields, via (\ref{currents}).  From the integrated equations of motion (\ref{EOM03}, \ref{EOM1}, \ref{EOM2}, \ref{EOM30}), we find the currents to be
\bea
d &=& c_0 \\
j_1 &=& c_1 + \frac{3}{2} h \mu  \\
j_2 &=& c_2 \\
j_3 &=& c_3 
\eea
where $\mu = a_0(\infty)$ is the baryon chemical potential and the other boundary values $a_i(\infty)$ have been chosen to be zero.  For notational convenience we define:
\bea
\label{tildeddef}
\tilde d &=& d + 3 h a_1 \\
\tilde j_1 &=& j_1 +  3h \left(a_0 - \frac{\mu}{2} \right) + 3 e a_3 \\
\tilde j_3 &=& j_3 - 3 e a_1 \ .
\eea
%In the tilde variables the integrated equations of motion are now
%\bea
%\label{EOM03}
%u^{5/2}\left(1+\frac{b^2}{u^3}\right)a_0' + \frac{eb}{u^3} a_3' &=&  \tilde d \sqrt{X} \\
%\label{EOM1}
%f(u) u^{5/2}\left(1+\frac{b^2}{u^3} -  \frac{e^2}{f(u)u^3}\right) a_1' &=&  \tilde j_1\sqrt{X}\\
%\label{EOM2}
%f(u) u^{5/2}a_2' &=&  j_2\sqrt{X}\\
%\label{EOM30}
%f(u) u^{5/2}(1- \frac{e^2}{f(u)u^3}) a_3' - \frac{eb}{u^3} a_0' &=&  \tilde j_3\sqrt{X} \ .
%\eea
We can rearrange (\ref{EOM03}) and (\ref{EOM30}), separating $a_0'$ and $a_3'$ as
\bea
\label{EOM0}
\frac{u^{5/2}\left(1+\frac{h^2}{u^3} -  \frac{e^2}{f(u)u^3}\right) a_0' }{\sqrt{X}}&=& \left(1-\frac{e^2}{f(u)u^3}\right) \tilde d - \frac{eh}{f(u)u^3}  \tilde j_3 \\
\label{EOM3}
\frac{u^{5/2}\left(1+\frac{h^2}{u^3} - \frac{e^2}{f(u)u^3}\right) f(u) a_3'}{\sqrt{X}} &=& \left(1+\frac{h^2}{u^3}\right) \tilde j_3 + \frac{eh}{u^3} \tilde d  \ .
\eea
We will first solve for $a_0'$ by dividing (\ref{EOM1}), (\ref{EOM2}), and (\ref{EOM3}) by (\ref{EOM0}) to give expressions for $a_1'$, $a_2'$, and $a_3'$ in terms of $a_0'$.  Plugging these back into (\ref{EOM0}) and solving for $a_0$ gives
\be
\label{soln0}
a_0 =\frac{\left(1-\frac{e^2}{f(u)u^3}\right) \tilde d - \frac{eh}{f(u)u^3}  \tilde j_3}{\sqrt{Z}}
\ee
where
\be
\label{Z}
Z = \left(1+\frac{h^2}{u^3} -  \frac{e^2}{f(u)u^3}\right)\left(u^5+\tilde d^2-\frac{j_2^2+\tilde j_3^2}{f(u)}\right) - \left(\frac{hd}{u^3}+\frac{e \tilde j_3}{f(u)u^3}\right)^2 - \frac{\tilde j_1^2}{f(u)} \ .
\ee
We can plug this back into the ratios of the equations of motion to find
\bea
\label{soln1}
f(u) a_1 &=& \frac{\tilde j_1}{\sqrt{Z}}\\
\label{soln2}
f(u) a_2 &=& \frac{\left(1+\frac{h^2}{u^3} - \frac{e^2}{f(u)u^3}\right)  j_2}{\sqrt{Z}} \\
\label{soln3}
f(u) a_3 &=& \frac{\left(1+\frac{h^2}{u^3}\right) \tilde j_3 + \frac{eh}{u^3} \tilde d}{\sqrt{Z}}  \ .
\eea

Using the solutions (\ref{soln0}, \ref{soln1}, \ref{soln2}, \ref{soln3}) we can write the DBI action (\ref{DBI}) in terms  of the currents as 
\be
S_{\rm DBI} = {\cal N} \int du \, u^{5/2} \frac{1+\frac{h^2}{u^3} -  \frac{e^2}{f(u)u^3}}{\sqrt{Z}} \ .
\ee
For the action to be real, $Z$ must clearly be non-negative.  The only way this can happen is if all three terms in (\ref{Z}) have double zeros at the same value of $u$, which we'll call $u_*$.  For the first term, this means each factor must vanish at $u_*$, so we get a total of four equations:
\bea
\label{one}
e^2 &=& f_* \left(u_*^3 + h^2 \right)  \\
\label{two}
j_{2*}^2 + \tilde j_{3*}^2 &=&  f_*\left(u^5_*+ \tilde d_*^2 \right) \\
\label{three}
f_* h\tilde d_* &=& - e \tilde j_{3*} \\
\label{four}
\tilde j_{1} &=& 0 \ .
\eea
To leading order in $e$, equation (\ref{one}) can be solved for $u_*$, giving
\be
u_*^3 = u_T^3 \left( 1+ \frac{e^2}{h^2 + u_T^3} \right)
\ee
which, again to leading order in $e$, then yields
\be
f_* = \frac{e^2}{h^2 + u_T^3} 
\ee
and $a_{1*} = a_{1T} + {\cal O}(e^2)$.  We can now solve (\ref{three}) for the Hall current:
\be
\label{Hallcurrent}
j_3 = \left( -\frac{h \tilde d_T}{u_T^3 + h^2} + 3a_{1T} \right) e \ . 
\ee
Plugging this result into (\ref{two}), we can solve for the ohmic current:
\be
\label{Ohmcurrent}
j_2 = \frac{\sqrt{u_T^8 + u_T^5 h^2 + u_T^3 \tilde d_T^2}}{u_T^3 + h^2} e \ .
\ee
Note that in the limit $h \to 0$ this reduces to the result (\ref{conductivityh=0}) found in \cite{BLL2}.  The final equation (\ref{four}) gives the axial current parallel to the magnetic field:
\be
j_1 = 3 h \left(\frac{\mu}{2} - a_{0*} \right) - 3 e a_{3*} \ .
\ee 
For small $e$, $a_{3} = {\cal O}(e)$ and $a_{0*} = a_{0T} + {\cal O}(e^2)$.  In addition, regularity at the horizon implies $a_{0T} = 0$, so we find 
\be
j_1 = \frac{3}{2} h \mu \ 
\ee 
which matches the result found in \cite{BLL3}.

\subsection{Conductivity by Kubo Formula}
\label{Kubo}

We will again calculate the ohmic and Hall conductivities, this time using the Kubo formula
\be
\label{Kuboformula}
\sigma_{ik} = \lim_{\omega \to 0} \frac{i}{\omega} < j_i j_k>_{\rm ret}(\omega)  
\ee
where the retarded current-current corelator is computed by
\be
< j_i j_k>_{\rm ret} = \frac{\delta S_{EOM}}{\delta a_i(\infty)\delta a_k(\infty)}
\ee
using a solution with infalling boundary conditions \cite{SonStarinets, PhotonDilepton, PS2}.

We begin with the background solution (background fields labeled with a bar) with a constant magnetic field $\b f_{23} = h$, discussed in \cite{BLL2}, consisting of $\b a_0$, $\b a_1$, and $\b a_3$ which obey the equations of motion
\bea
\label{EOMbA0}
\frac{u^{5/2} \left(1+\frac{h^2}{u^3 }\right) \b a_0'}{\sqrt C} &=& d + 3h \b a_1 \\
\label{EOMbA1}
\frac{u^{5/2} \left(1+\frac{h^2}{u^3}\right) f(u) \b a_1'}{\sqrt C} &=& j_1 + 3h \left( \b a_0 - \frac{\mu}{2}\right) \\
\b a_3 &=& h x_2
\eea
where $d$ is the total baryon charge, $\mu$ is the baryon chemical potential, $j_1 = \frac{3}{2} h\mu$ is the axial current generated by the magnetic field, and  
\be
C= \left(1+\frac{h^2}{u^3}\right) \left(1- {\b a_0'}^2 + f(u) {\b a_1'}^2 \right) \, .
\ee

With an eye toward an electric field and currents in the 2- and 3-directions, we consider perturbations to the gauge field of the form $a_{2/3}(u, t) = e^{-i\omega t}\alpha_{2/3}(u)$, which decouple from perturbations in $a_0$ and $a_1$.  The action, to second order in the perturbations, consists of a DBI term,
\be
S_{DBI} ={\cal N} \int \frac{du \, u^{5/2}}{2\sqrt C}\left( - \frac{1}{f(u) u^3} \left({\dot a_2}^2+{\dot a_3}^2\right) + f(u)\left( {a_2'}^2+ {a_3'}^2\right) - 2 \frac{\b a_0' h}{u^3} \left(\dot a_2 a_3' - a_2' \dot a_3\right) \right)
\ee
and a Chern-Simons term,
\be
S_{CS}=  {\cal N} \int \left( \b a_1'  a_{[2} \dot a_{3]} - \b a_1 a_{[2}' \dot a_{3]}  \right)
\ee

Using $e^{-i\omega t}$ as an ansatz for the time dependence, the equations of motion are as follows:
\bea
\label{EOMA2}
\partial_u \left( \frac{f(u)u^{5/2} \alpha_2'}{\sqrt C} \right) + \omega^2 \frac{\alpha_2}{f(u)u^{1/2} \sqrt C} & = &  -i\omega \left(-\partial_u \left(\frac{\b a_0' h}{u^{1/2}\sqrt{C}}\right) + 3 \b a_1' \right) \alpha_3 \\
\label{EOMA3}
\partial_u \left( \frac{f(u)u^{5/2} \alpha_3'}{\sqrt C} \right) + \omega^2 \frac{\alpha_3}{f(u)u^{1/2} \sqrt C} & = &  i\omega \left(-\partial_u \left(\frac{\b a_0' h}{u^{1/2}\sqrt{C}}\right) + 3 \b a_1' \right) \alpha_2 \ .
\eea
  
Changing to the variable $y = 2u^{-1/2}$ (and using $\frac{\partial y}{\partial u} = -y^3/8$), we can write (\ref{EOMA2}) and (\ref{EOMA3}) as
\be
\label{EOMAperp}
\partial_y \left( \frac{f(u) \partial_y\alpha_{2/3}}{y^2 \sqrt C} \right) + \frac{\omega^2}{f(u)y^2 \sqrt C} \alpha_{2/3} \pm \frac{i\omega}{4} \partial_y \gamma \alpha_{3/2}=0
\ee
where $\gamma = -\frac{y^4 h \partial_y \b a_0 }{16\sqrt{C}} - 3 \b a_1$ which is function only of the background solution.  

As the Kubo formula requires us to compute the retarded propagator, we need to pick out the solutions with infalling boundary conditions at the horizon, $y=y_T$.  Defining $z=y_T-y$, the near-horizon limit of (\ref{EOMAperp}) is
\be
\label{EOMnearhorizon}
\partial_y^2 \alpha_{2/3} - \frac{1}{z} \partial_y \alpha_{2/3} + \frac{\omega^2 y_T^2}{36 z^2}  \alpha_{2/3} = 0
\ee
which comes only from the first two term of (\ref{EOMAperp}); the third term of (\ref{EOMAperp}) dropped out as it was higher order in $z$.  The solution to (\ref{EOMnearhorizon}) is
\be
\alpha_{2/3} = z^{\pm i \nu} 
\ee
where the frequency $\nu = \frac{\omega y_T}{6}$.  For the infalling solution, we choose the minus sign: $\alpha_{2/3} = z^{- i \nu} $.

For general $z$, we now expand $\alpha_{2/3}$ in powers of $\nu$: 
\bea
\alpha_{2/3} &=& z^{-i \nu} \left(G^{(0)}_{2/3} + \nu G^{(1)}_{2/3} + {\cal O}(\nu^2) \right) \\
&=& (1 - i\nu \log z) G^{(0)}_{2/3} + \nu G^{(1)}_{2/3} + {\cal O}(\nu^2) \ . 
\eea
We need to solve for the expansion coefficients $G^{(0)}_{2/3}$ and  $G^{(1)}_{2/3}$.  To zeroth order in $\nu$, the equation of motion (\ref{EOMAperp}) is just
\be
\partial_y \left( \frac{f(u)}{y^2 \sqrt C} \partial_y G^{(0)}_{2/3} \right) = 0
\ee
which has two solutions, one where $G^{(0)}_{2/3}$ is constant and one where it goes as $\log z$ near the horizon.  The constant solution has the correct boundary conditions, so we keep it and discard the $\log z$ solution.  

Now to first order in $\nu$, (\ref{EOMAperp}) can be integrated once to give
\be
\label{G}
\partial_y G^{(1)}_{2/3} = \frac{-iG^{(0)}_{2/3}}{z} + \frac{k_1 y^2 \sqrt{C}}{f(u)}  \mp \frac{3iy^2\sqrt{C}}{2y_T f(u)} \gamma G^{(0)}_{3/2}
\ee
where $k_1$ is a constant of integration.  We can integrate (\ref{G}) in the near-horizon limit to yield 
\be
G^{(1)}_{2/3} = \left( iG^{(0)}_{2/3} - \frac{k_1 y_T^3 \sqrt{C_T}}{6}  \pm \frac{iy_T^2\sqrt{C_T}}{4} \gamma_T G^{(0)}_{3/2} \right) \log z + k_2
\ee
where $k_2$ is another integration constant. Since the infalling boundary condition forbids a $\log z$ divergence at the horizon, the coefficient in the parenthesis must be zero, which implies
\be
k_1 = \frac{6i G^{(0)}_{2/3}}{y_T^3 \sqrt{C_T}} \pm \frac{3i}{2 y_T} \gamma_T G^{(0)}_{3/2}
\ee
We can now solve for $G^{(1)}_{2/3}$ near the boundary.  Using $\partial_y \b a_0 \to -\frac{y^2d}{4}$ as $y \to 0$, we integrate (\ref{G}) to obtain
\bea
G^{(1)}_{2/3} &=& k_2 + i G^{(0)}_{2/3} \log z + \frac{k_1 y^3}{3} \pm \frac{i y^9 h d}{384 y_T}G^{(0)}_{3/2}\\
&=& k_2 + i G^{(0)}_{2/3} \log z+ i \left( \frac{2}{\sqrt{C_T}} G^{(0)}_{2/3} \pm \frac{y_T^2 \gamma_T}{2} G^{(0)}_{3/2} \right) \left(\frac{y}{y_T}\right)^3 + {\cal O}(y^9)
\eea
Reassembling $\alpha_{2/3}$, we see that the $\log z$ terms cancel.  So, to first order in $\omega$,
\be
a_{2/3} = G^{(0)}_{2/3} +\frac{\omega y_T k_2}{6}  - i\omega t G^{(0)}_{2/3} +  \frac{i \omega y_T}{6} \left( \frac{2}{\sqrt{C_T}} G^{(0)}_{2/3} \pm \frac{y_T^2 \gamma_T}{2} G^{(0)}_{3/2} \right) \left(\frac{y}{y_T}\right)^3 \ .
\ee
Since $e_{2/3} = \dot a_{2/3}$, we can relate it to the constant $G^{(0)}_{2/3}$, to leading order in $\omega$, as
\be
e_{2/3} =  -i\omega G^{(0)}_{2/3} \ .
\ee

Having computed $a_{2/3}$, we readily find the current, to leading order in $\omega$, to be
\be
j_{2/3} = \frac{\delta S_{EOM}}{\delta a_{2/3}(\infty)} ={\frac{\delta L}{\delta a'_{2/3}} \vline}_{u \to \infty}  = -i\omega\left( \frac{u_T G^{(0)}_{2/3}}{\sqrt{C_T}} \pm \gamma_T G^{(0)}_{3/2}  \right)
\ee
where we identify the first term as the ohmic current and the second term as the Hall current. The ohmic conductance in the 2-direction is then given via the Kubo formula (\ref{Kuboformula}) as
\be
\sigma_{22} =  \lim_{\omega \to 0} \frac{i}{\omega} \frac{\delta^2 S_{EOM}}{\delta a_2(\infty) \delta a_2(\infty)} = \frac{u_T}{\sqrt{C_T}} \ . 
\ee
Using the equation of motion (\ref{EOMbA0}), we solve for $\b a_{0T}$ in terms of $\tilde d_T$ which allows us to rewrite $C_T$ as\footnote{Here we define $\tilde d_T = d + 3h \b a_{1T}$ which, strictly speaking, is not the $\tilde d$ defined previously in (\ref{tildeddef}).  However, since we are working to lowest order in $e$, the two definitions are equivalent.}
\be
C_T = \frac{ \left(u_T^3+ h^2\right)^2}{ u_T^6 + u_T^3 h^2 + u_T {\tilde d_T}^2} \, .
\ee
From this we find
\be
\sigma_{22} =  \frac{\sqrt{ u_T^8 + u_T^5 h^2 + u_T^3 {\tilde d_T}^2}}{u_T^3+ h^2} \ .
\ee
which is the same as the result (\ref{Ohmcurrent}) from the previous section using the Karch-O'Bannon method.  An analogous calculation also gives $\sigma_{33} = \sigma_{22}$.

We can also compute, using the Kubo formula (\ref{Kuboformula}), the Hall conductivity:
\bea
\sigma_{32} =  \lim_{\omega \to 0} \frac{i}{\omega} \frac{\delta^2 S_{EOM}}{\delta a_3(\infty) \delta a_2(\infty)}  &=& - \gamma_T \\
&=& -\frac{h(\partial_u \b a_0)|_T}{\sqrt{u_T C_T}} + 3 \b a_{1T}  \ .
\eea
We again use the equation of motion (\ref{EOMbA0}) for $\b a_0$ to find
\be
\sigma_{32} =  -\frac{h \tilde d_T}{u_T^3 +h^2} + 3 \b a_{1T}  \ ,
\ee
and the result again exactly matches the Karch-O'Bannon result (\ref{Hallcurrent}).   Repeating the calculation with indices exchanged gives $\sigma_{23} = - \sigma_{32}$.

\subsection{Possible Interpretation}

\begin{figure}
\centerline{\epsfig{file= 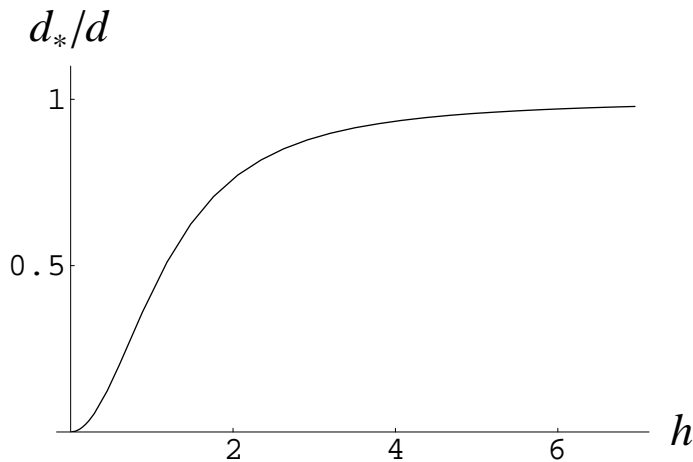,width=10cm}}
\caption{The ratio of the charge $d_*$ to the total charge $d$ as a function of the magnetic field $h$ at temperature $T=0.3$.}
\label{dstarfig}
\end{figure}

Let us try to interpret these results.  Since $a_{1}(\infty) =0$ and is decreasing towards the horizon, we can define a positive charge $d_{*}=-3h a_{1T}$.  The generic behavior\footnote{At low temperature or at large densities there can be a phase transition \cite{LL2}.}  of $d_{*}$ is shown in figure \ref{dstarfig}.

We would like to interpret $d_{*}$ as the charge which is located in the bulk above the horizon as smeared 4-branes inside the 8-branes.  The Hall conductivity can be written as 
\begin{equation}
\sigma_{23}=-\frac{h \tilde d}{h^2 +u_{T}^{3}}-\frac{d_{*}}{h}
\end{equation}
This form looks like we have two kinds of charge carriers in the system. The first, whose charge is $\tilde d = d-d_{*}$,  behaves like charges flowing through a neutral dissipative media, which, in the holographic QCD interpretation of the model, are the gluons.  The second type, whose charge is $d_{*}$, seems 
to behave as if it does not interact with the dissipative bath at all.  This interpretation is consistent with the result for the DC longitudinal conductivity where only the first kind of charge contributes.  Indeed, charges without dissipation can not contribute to the DC conductivity.

We can relate these properties to the bulk supergravity point of view as follows.  The ordinary charge carriers are chiral fermions, i.e. quarks, which correspond in the bulk to short strings stretching from the 8-brane to the horizon.  The horizon, in a rough sense, represents the interactions with the $SU(N_c)$ degrees of freedom in the thermal vacuum.  Baryons are singlets under the $SU(N_c)$ and, as a result, do not interact with the gluonic bath.  From the bulk point of view, a baryon corresponds to $N_c$ strings stretching from the 8-brane to a wrapped 4-brane, not to the horizon.  The smeared 4-branes, likewise, are up above the horizon and have $N_c$ strings connecting them to the 8-branes.  Correspondingly, the charge carriers they represent do not feel the dissipative thermal degrees of freedom.  However, the field theory interpretation of why some of the charge is dissipationless is unclear.

\section{Parallel electric and magnetic fields}
\label{parallel}

We now turn to the case of parallel electric and magnetic fields.  Here one expects from the four-dimensional point of view that the axial current will no longer be conserved as a consequence of the axial anomaly.  In particular, if we consider a magnetic field $f_{23}=\partial_2 a_3 = h$ and electric field $f_{01} =\dot a_1 = e$, the right hand side of the anomaly equation
\begin{equation}
\partial_{\mu}j^{\mu}_{A} \sim f \wedge f \sim eh
\label{4danom}
\end{equation}
is now nonzero. 

In section \ref{orthogonal} we had an ordinary vector charge density $d \equiv j_{0V}$ and an anomally-generated axial current $j_{1A}$.  Here, the situation is reversed; the electric field will generate a vector current $j_{1V}$ and due to the anomally and the magnetic field will generate axial charge $d_A \equiv j_{0A}$.  For a recent discussion in the context of QCD, see \cite{FKW}.  In the bulk, we will therefore consider the corresponding components of the 8-brane and $\b 8$-brane gauge fields, namely that $a_0$ is axial and $a_1$ is vector.  That is, $a_0^{D8} = -a_0^{\b{D8}}$ and $a_1^{D8} = a_1^{\b{D8}}$.
 
The anomaly arises in the bulk description through the non-gauge invariance of the Chern-Simon term of the 8-brane action under gauge transformations that do not vanish as $u \rightarrow \infty$ \cite{SS}.   This is illustrated by the fact that parallel electric and magnetic fields lead to a source in the $a_{u}$ equation of motion.  For orthogonal $e$ and $h$, we had assumed that the currents were homogenous and stationary, so that $a_0'$ and $ a_1'$ depended only on $u$ and not space or time.  However, now we cannot both use this ansatz and the $a_{u}=0$ gauge because then the $a_u$ equation of motion would not be satisfied.  Since this gauge is very useful, we still want to work with it.  Let us therefore relax our ansatz and allow $a_\mu'$ to depend on $x_\mu$.  In the simplest case, where only $a_0'$ depends on time, the equation of motion for $a_{u}$ can be satisfied with $a_{u}=0$ provided
\begin{equation}
\partial_{t}\left(\frac{u^{5/2}(1+\frac{h^2}{u^{3}})a_{0}^{'}}{\sqrt{C}}\right)=3eh
\label{auconst}
\end{equation}
 which is the holographic analog of (\ref{4danom}).  Taking $u \rightarrow \infty$, this becomes
\begin{equation}
\frac{d}{dt} d_{A}=\frac{3}{2}eh \ .
\end{equation}
As expected, once we consider an electric field parallel to a magnetic field the axial charge is not conserved.  

Let us imagine we start with a system with no axial charge but with a magnetic field turned on.  We now turn on an electric field parallel to the magnetic field; axial charge is produced and increases with time. The existence of an axial charge means that the axial chemical potential $\mu_{A}$ is now nonzero and also increases with time. As explained in \cite{BLL3}, this means that, again because of the anomaly, there is a parallel vector current $j_{1V} =  \frac{3}{2}h \mu_{A}$ increasing in time.  Indeed, as we will see, the conductivity at zero frequency will have a delta function indicating a perfect conductor.

The computation of the conductivity will follow the Kubo formula as in section \ref{Kubo}.  Again, we denote by $\b a_\mu$ the background gauge fields with $e=0$ (note that $\b a_0$ is a vector and $\b a_1$ is axial), and we consider time-dependent perturbations $a_\mu$.  The equations in this case involve only $a_{0}$ and $a_{1}$, 

\begin{eqnarray}
\label{sysper}
\partial_{u}\left(\frac{u^{5/2}(1+\frac{h^2}{u^{3}})a_{0}^{'}}{\sqrt{C}}\right) &=& 3ha_{1}^{'} \\
\partial_{u}\left(\frac{f(u)u^{5/2}(1+\frac{h^2}{u^{3}})a_{1}^{'}}{\sqrt{C}}\right)+\omega^{2}\frac{(1+\frac{h^2}{u^{3}})a_{1}}{f(u)u^{1/2}\sqrt{C}} &=& 3ha_{0}^{'} \ ,
\end{eqnarray}
together with the $a_u$ equation of motion (\ref{auconst}) which acts as a constraint on the perturbations in the fields.  However, this constraint just implies that in the integrated form of (\ref{sysper}) no time-dependent constant of integration is allowed:
\be
\frac{u^{5/2}(1+\frac{h^2}{u^{3}})a_{0}^{'}}{\sqrt{C}} = 3ha_{1} \ .
\ee
Substituting in for $a_0' $, the equation for $a_{1}$ then takes the form
\begin{equation}
\label{a1withmass}
\partial_{u}\left(\frac{f(u)u^{5/2}(1+\frac{h^2}{u^{3}})a_{1}^{'}}{\sqrt{C}}\right)+\omega^{2}\frac{(1+\frac{h^2}{u^{3}})a_{1}}{f(u)u^{1/2}\sqrt{C}}=\frac{9h^2\sqrt{C}}{u^{5/2}(1+\frac{h^2}{u^{3}})}a_{1} \ .
\end{equation}
We see from this equation that the field $a_{1}$ looks like it has acquired a mass; we thus expect the conductivity to have a delta function singularity at zero frequency.

In the small $h$ approximation and for $\omega = 0$, one can solve the system analytically.  At lowest order in $h$, (\ref{EOMbA1}) gives $\bar{a}_{1}=0$ and from  (\ref{a1withmass}) $a_{1}$ is a constant: $a_{1}=a_{1}(\infty)$.  We thus find using (\ref{a1withmass}) that
\bea
j_{1V} = \frac{\delta S_{EOM}}{\delta a_1(\infty)}&=& \left. \frac{f(u) u^{5/2} a_1'}{\sqrt{1-{\b a_0'}^2}} \right|_{u \to \infty} - \frac{3}{2} h a_1(\infty) \nonumber \\
&=&\frac{9}{2}h^2 a_{1}(\infty) \int_{u_{T}}^{\infty}\frac{\sqrt{1-{\b a_0'}^2}}{u^{5/2}} \ .
\eea
To zeroth order in $h$, the integrated equation of motion for $\b a_0$ (\ref{EOMbA0}) can be rewritten as
\begin{equation}
\b a_{0}^{'}=\frac{d}{\sqrt{u^5 +d^2}} \ ,
\end{equation}
so now
\be
j_{1V} = \frac{9h^2\mu}{2d}  a_1(\infty) 
\ee
where $\mu=a_{0}(\infty)$ is the baryon chemical potential.  The two-point vector current correlator at zero frequency to lowest order in $h$ is therefore
\begin{equation}
 \left. <j_{1}j_{1}> \right|_{\omega =0} = \left. \frac{\delta^2 S_{EOM}}{\delta a_1(\infty)\delta a_1(\infty)} \right|_{\omega=0} = \frac{9h^2\mu}{2d}  \ .
\end{equation}

Because the correlator is non-vanishing at $\omega = 0$, the Kubo formula (\ref{Kuboformula}) implies that there is a pole in the imaginary part of the conductivity at $\omega \rightarrow 0$.
Since the real and imaginary parts of the conductivity obey
\begin{equation} 
{\rm Im} [\sigma(\omega)]=-\frac{1}{\pi}{\cal P}\int_{-\infty}^{\infty}\frac{\Re [\sigma(\omega^{'})]}{\omega^{'}-\omega}d\omega^{'} \ ,
\end{equation}
we infer that the real part of the conductivity has a delta function.  In particular, for small $h$ and as $\omega \rightarrow 0$,
\begin{equation}
{\rm Re} [\sigma(\omega)]\sim \frac{9\pi h^2\mu}{2d} \delta (\omega) 
\end{equation}
which implies that for $e$ parallel to $h$ the system is a perfect conductor even in its vacuum state.  Note that the coefficient of the delta function diverges when both $T$ and $d \to 0$.  At large $d$, the coefficient falls off as $d^{-3/5}$, while for $d=0$, it is proportional to $T^{-3}$.

\section*{Acknowledgments}
We wish to thank Oren Bergman, Niko Jokela, and Ady Stern for useful discussions.
This work was supported in part by the Israel Science Foundation under grant no.~568/05.

\end{document}